\documentclass[preprint,aps,amsmath,superscriptaddress,nofootinbib]{revtex4}
\usepackage{bm}
\usepackage{epsfig}

\usepackage{ifpdf}

\newcommand{\nn}{\nonumber}
\newcommand{\bqa}{\begin{eqnarray}}
\newcommand{\eqa}{\end{eqnarray}}

\begin{document}

\ifpdf
\DeclareGraphicsExtensions{.pdf, .jpg}
\newcommand{\picspace}{\vspace{-2.5in}}
\newcommand{\picspacehalf}{\vspace{-1.75in}}
\else
\DeclareGraphicsExtensions{.eps, .jpg,.ps}
\newcommand{\picspace}{\vspace{0in}}
\newcommand{\picspacehalf}{\vspace{0in}}
\fi


\preprint{\vbox{ \hbox{GUCAS-CPS-06-12}   }}

\title{Hunting $\eta_b$ through radiative decay into $J/\psi$}

\author{Gang Hao\footnote{Electronic address: hao$_-$gang@mails.gucas.ac.cn}}
\affiliation{Department of Physics, Graduate University of Chinese
Academy of Sciences,
YuQuan Road 19A, Beijing 100049, China\vspace{0.2cm}}

\author{Yu Jia\footnote{Electronic address: jiay@ihep.ac.cn}}
\affiliation{Institute of High Energy Physics, Chinese Academy of Sciences,
Beijing 100049, China\vspace{0.2cm}}

\author{Cong-Feng Qiao\footnote{Electronic address: qiaocf@gucas.ac.cn}}
\affiliation{CCAST (World Lab.), P.O. Box 8730, Beijing 100080,
China\vspace{0.2cm}} \affiliation{Department of Physics, Graduate
University of Chinese Academy of Sciences, YuQuan Road 19A,
Beijing 100049, China\vspace{0.2cm}}

\author{Peng Sun\footnote{Electronic address: sunpeng05@mails.gucas.ac.cn}}
\affiliation{Department of Physics, Graduate University of Chinese
Academy of Sciences,
YuQuan Road 19A, Beijing 100049, China\vspace{0.2cm}}

\date{\today\\ \vspace{1cm} }



\begin{abstract}

We propose that the radiative decay process, $\eta_b\to
J/\psi\,\gamma$,  may serve as a clean searching mode for $\eta_b$
in hadron collision facilities. By a perturbative QCD calculation,
we estimate the corresponding branching ratio to be of order
$10^{-7}$. Though very suppressed, this radiative decay channel in
fact has larger branching ratio than the hadronic decay process
$\eta_b\to J/\psi\,J/\psi$, which was previously hoped to be a
viable mode to search for $\eta_b$ in Tevatron Run 2. The
discovery potential of $\eta_b$ through this channel seems
promising in the forthcoming LHC experiments, and maybe even in
Tevatron Run 2,  thanks to the huge statistics of $\eta_b$  to be
accumulated in these experiments. The same calculational scheme is
also used to estimate the branching ratios for the processes
$\eta_b\,(\eta_c) \to \phi\gamma$.

\end{abstract}

\maketitle

\newpage

\section{Introduction}

The existence of $\eta_b$, the pseudo-scalar partner of
$\Upsilon(1S)$, is a firm prediction of QCD, about which nobody
would seriously challenge.
It is rather unsatisfactory that although three decades
have elapsed since the discovery of $\Upsilon(1S)$,
this particle eludes intensive experimental endeavors
and still eagerly awaits to be established.

On the theoretical side, many work have been devoted to uncovering
various properties of $\eta_b$, such as its mass, inclusive
hadronic and electromagnetic widths, transition rates and
production cross sections in different collider
programs~\cite{Brambilla:2004wf}. Among all the observables, its
mass is believed to be the simplest and most unambiguous to
predict.  Recent estimates for $\Upsilon-\eta_b$ mass splitting
span the 40--60 MeV range~\cite{Ebert:2002pp,
Recksiegel:2003fm,Kniehl:2003ap, Gray:2005ur}. An eventual
definite sighting of $\eta_b$ and precise measurement of its mass
will critically differentiate varieties of theoretical approaches,
consequently sharpening our understanding towards the $b\bar{b}$
ground state.

The $\eta_b$ has  recently been sought from $\gamma\gamma$
collisions in the full LEP 2 data sample,  where approximately few
hundreds of $\eta_b$ are expected to be produced. ALEPH has one
candidate event with the reconstructed mass of $9.30\pm 0.03$ GeV,
but consistent to be a background event~\cite{Heister:2002if}.
ALEPH, L3, DELPHI have also set upper limits on the branching
fractions for $\eta_b$ decays into 4, 6, 8 charged
particles~\cite{Heister:2002if,Levtchenko:2004ku,Abdallah:2006yg}.
Based on the 2.4 ${\rm fb}^{-1}$ data taken at the $\Upsilon(2S)$
and $\Upsilon(3S)$ resonances, CLEO has searched distinctive
single photons from hindered $M1$ transitions
$\Upsilon(2S),\Upsilon(3S)\to \eta_b\gamma$, and from the cascade
decay $\Upsilon(3S)\to h_b\pi^0,\:h_b\pi^+\pi^-$ followed by $E1$
transition $h_b\to \eta_b\gamma$, but no signals have been
found~\cite{Artuso:2004fp}.

Hadron collider experiments provide an alternative means  to
search for $\eta_b$. Unlike the $e^+e^-$ machines which are limited
by the low yield of $\eta_b$,  hadron colliders generally
possess a much larger $\eta_b$ production rate, which in
turn allows for triggering it through some relatively rare decay
modes yet with clean signature.  However, one caveat is also worth
being emphasized.  The corresponding background events
are in general copious in hadron machines,
so the virtue of this sort of decay modes may be
seriously discounted (Such an example is the
electromagnetic decay $\eta_b\to \gamma\gamma$,
with an expected branching fraction $\sim 10^{-4}$,
which is nevertheless overshadowed by the
ubiquitous $\gamma$ events
originating from $\pi^0$ decay).

Several years ago, Braaten, Fleming and Leibovich
suggested that the hadronic decay $\eta_b\to J/\psi J/\psi$,
followed by both $J/\psi$ decays to muon pairs,
can be used as a very clean trigger to search for
$\eta_b$ at Tevatron Run 2~\cite{Braaten:2000cm}.
By some simple scaling assumption,
they estimate the branching ratio of the double
$J/\psi$ mode to be $7\times 10^{-4\pm 1}$, and conclude
that the prospect of observing the $\eta_b \to 4\mu$ channel
at Tevatron Run 2 is rather promising.
CDF in fact has followed this suggestion and looked
at the full Run 1 data for the $4\mu$ events
in the expected $\eta_b$ mass window~\cite{Tseng:2003md}.

However, some objection has been recently put forward
by Maltoni and Polosa, who argue that the estimate of
the branching ratio of double $J/\psi$ mode by Braaten
{\it et al} might be too optimistic~\cite{Maltoni:2004hv}.
They instead advocate that $\eta_b\to D^* \overline{D}^{(*)}$,
with the estimated decay ratios lying in the
$10^{-3}-10^{-2}$ range, may serve as better searching
modes for $\eta_b$ in Run 2.

Very recently, one of us (Y.~J.) has surveyed the discovery
potential of various hadronic decay channels of
$\eta_b$~\cite{Jia:2006rx}. The explicit perturbative QCD (pQCD)
calculation predicts ${\rm Br}[\eta_b\to J/\psi J/\psi]$ to be
only of order $10^{-8}$. If this is the case, the chance of
observing this decay mode in Run 2 then becomes rather gloomy,
whereas the observation possibility at LHC may still remain.
Another noteworthy assertion of \cite{Jia:2006rx} is that, by some
rough but physical considerations,  one expects ${\rm
Br}[\eta_b\to D^*\overline{D}]\sim 10^{-5}$ and ${\rm
Br}[\eta_b\to D^*\overline{D}^*]\sim 10^{-8}$, which are much
smaller than the optimistic estimates by Maltoni and Polosa.
Taking the low reconstruction efficiency of $D$ mesons further
into consideration, these double charmful decay modes may not be
as attractive as naively thought.

In this paper, we try to propose another decay process, $\eta_b\to
J/\psi \gamma$, as a viable discovery channel for $\eta_b$ in
hadron collider experiments. At first sight, this mode, being a
radiative decay process, may not look very economical due to the
large total width of $\eta_b$. In fact, our explicit pQCD
calculation reveals that the corresponding branching ratio is
indeed very suppressed, only about $10^{-7}$. Nevertheless, it is
worth emphasizing that this is already {\it larger} than that of
the double $J/\psi$ mode. The usual way of reconstructing
$J/\psi$ is via the dimuon mode, with a branching fraction about
6\%. In the light of this, the advantage of this radiative decay
process over the double $J/\psi$ mode will be further amplified,
since only one $J/\psi$ needs to be reconstructed in our case.
Unlike the open-charm decay processes $\eta_b\to
D^*\overline{D}^{(*)}$, in which the final decay products such as
$K$, $\pi$ in general suffer severe contamination from the
combinatorial background, the presence of $J/\psi$ in the final
state renders our radiative decay channel much cleaner to look for
experimentally.

Thanks to the huge amount of $\eta_b$  to be accumulated in high
energy hadron collision facilities, we expect that this radiative
decay channel, albeit being a rare decay mode, may still have
bright prospect to be observed in Run 2 and in the forthcoming LHC
program.  Not surprisingly, one should be aware that the major QCD
background events, {\it i.e.} the associated $J/\psi+\gamma$
production~\cite{Roy:1994vb,Kim:1996bb,Mathews:1999ye}, may
exceedingly outnumber our signal events. To make this mode
practically viable, one has to ensure that those background events
can be significantly depressed  by judiciously adjusting the
kinematical cuts.

The remainder of the paper is distributed as follows. In
Sec.~\ref{CSM:Calculation}, we present the pQCD calculation for
the radiative decay process $\eta_b\to J/\psi\gamma$, treating
heavy quarkonium states in NRQCD approach which, to our purpose,
is equivalent to the color-singlet model. In
Sec.~\ref{phenomenology}, we present the numerical prediction to
the corresponding branching ratio,  and explore the observation
potential of this decay mode in Tevatron  Run 2 and in the coming
LHC experiment. We subsequently apply the same formalism to
estimate the branching ratios for analogous processes
$\eta_b\,(\eta_c) \to \phi \gamma$,  which is equivalent to work
in the context of constituent quark model. We summarize in
Sec.~\ref{summary}.

\section{Color-singlet Model Calculation}
\label{CSM:Calculation}

\begin{figure}[tb]
\begin{center}
\includegraphics[scale=0.9]{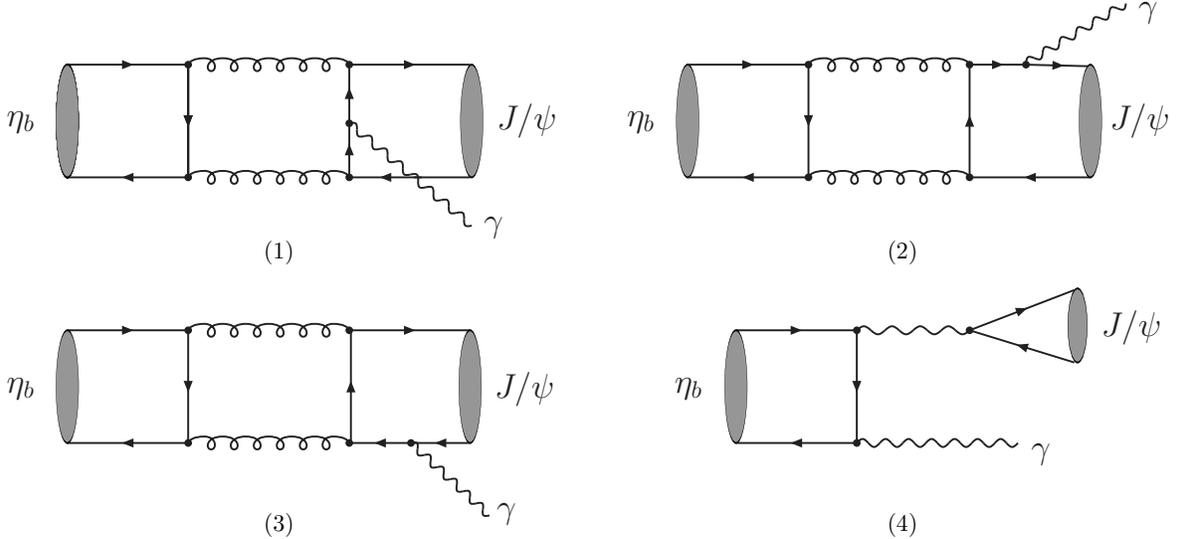}
\caption{Lowest-order diagrams that contribute to
$\eta_b\rightarrow J/\psi\,\gamma$.  Diagrams (1), (2), (3) stand
for QCD-initiated  process, while (4) represents the dominant QED
process governed by photon fragmentation. Crossed diagrams are
implicitly implied. \label{Feynman:Diag}}
\end{center}
\end{figure}

In this section, we present a pQCD calculation for the decay rate
of $\eta_b\to J/\psi \gamma$. This decay process can be initiated
by either strong or electromagnetic interaction, with the
corresponding lowest-order diagrams shown in
Fig.~\ref{Feynman:Diag}. Since the annihilation of the $b\bar{b}$
pair, the creation of the $c \bar{c}$ pair, as well as the
emission of the hard photon, are all dictated by short-distance
physics, with the hard scales set by the heavy quark masses, it is
appropriate to utilize the pQCD scheme to tackle this exclusive
process.

Nowadays it has become standard to employ the NRQCD factorization
framework to cope with hard processes involving heavy
quarkonia~\cite{Bodwin:1994jh}.  One of the major advantages of
this model-independent EFT approach over the old color-singlet
model is that it can in principle incorporate the contribution of
higher Fock components of quarkonium ({\it color-octet effect}) in
a systematic fashion. This is exemplified by the systematic
treatment of inclusive annihilation decay of various quarkonium
states within this framework~\cite{Bodwin:1994jh,Brambilla:2002nu}. 
In contrast, the inclusion of color-octet effect in exclusive
processes has not been developed to a comparable level within this
context.  To date, there are only few cases where this effect has
been investigated in the model-independent language.  One example
is the magnetic dipole transition process such as $J/\psi\to
\eta_c\gamma$, where the leading color-octet contribution
vanishes~\cite{Brambilla:2005zw}. Another example is the
quarkonium radiative decay to light mesons, such as $\Upsilon \to
f_2(1270)\gamma$, where the color-octet effect turns out to be
quite insignificant~\cite{Fleming:2004hc}.

If we take the lessons from the aforementioned radiative decay
processes, it seems persuasive to assume that the color-octet
effect may also be unimportant in our case. We will completely
ignore this effect in this work. In this regard, there will be no
difference between the NRQCD approach and the color-singlet model,
consequently these two terms will be used interchangeably.

Before launching into the actual calculation, it is worth
mentioning that,  Guberina and Kuhn has studied the analogous
radiative decay process $\Upsilon\to \eta_c\gamma$ in NRQCD
approach more than two decades ago~\cite{Guberina:1980xb}. Our
calculation will intimately resemble theirs.

It is useful to take notice of a simple trait of this process,
that the photon can only be transversely polarized, so is the
recoiling $J/\psi$ by angular momentum conservation. This property
holds true irrespective of whether this process is initiated by
strong or electromagnetic interaction. In fact, parity and Lorentz
invariance constrains the decay amplitude to have the following
unique tensor structure:
\bqa {\cal M}(\lambda_1, \lambda_2) &=& {\cal A} \:\epsilon_{\mu
\nu \alpha \beta}\, \varepsilon^{*\mu}_{J/\psi}(\lambda_1)\,
\varepsilon^{*\nu}_{\gamma}(\lambda_2)\,Q^\alpha\,k^\beta\,.
\label{Lorentz:tensor:structure} \eqa
We  use $Q$, $P$ and $k$ to signify the momenta of $\eta_b$,
$J/\psi$ and $\gamma$, respectively, and use $\lambda_1$ and
$\lambda_2$ to label the helicities of $J/\psi$ and $\gamma$ which
are viewed in the $\eta_b$ rest frame. It is straightforward to
infer from (\ref{Lorentz:tensor:structure}),  that the only
physically allowed helicity configurations are ($\lambda_1$,
$\lambda_2$)=($\pm 1$, $\pm 1$). All the dynamics is encoded in
the coefficient ${\cal A}$, which we call {\it reduced} amplitude.
Our task then is to find out its explicit form.

In the color-singlet model, it is customary to start with the
parton process $b(p_b)\,\bar{b}(p_{\bar b}) \to c(p_c)
\,\bar{c}(p_{\bar{c}}) + \gamma(k)$, then project this matrix
element onto the corresponding color-singlet quarkonium Fock
states. For reactions involving heavy quarkonium, it is
conventional to organize the amplitude in powers of the relative
momentum between its constituents, to accommodate the NRQCD
ansatz. This work is intended for the leading order accuracy in
relativistic expansion only, it is then legitimate to neglect the
relative momentum inside both $\eta_b$ and $J/\psi$.  We thus set
$p_b=p_{\overline b}=Q/2$ and $p_c=p_{\overline c}=P/2$. For the
$b \overline{b}$ pair to form $\eta_b$, it is necessarily in a
spin-singlet and color-singlet state, and one can replace the
product of the Dirac and color spinors for $b$ and $\overline{b}$
in the initial state with the projector
\bqa
u(p_b)\,\overline{v}(p_{\bar b})& \longrightarrow& {1\over 2
\sqrt{2}} \,(\not\! Q+ 2 m_b)\,i\gamma_5 \times \left( {1\over
\sqrt{m_b}} \psi_{\eta_b}(0)\right) \otimes \left( {{\bf 1}_c\over
\sqrt{N_c}}\right)\,.
\label{Etab:projector}
\eqa
For the outgoing $J/\psi$, one can employ the following
projection operator:
\bqa
v(p_{\bar c})\,\overline{u}(p_c)& \longrightarrow& {1\over 2
\sqrt{2}} \not\! \epsilon^*_{J/\psi}\,(\not\! P+2 m_c)\, \times
\left( {1\over \sqrt{m_c}} \psi_{J/\psi}(0)\right) \otimes \left(
{{\bf 1}_c\over \sqrt{N_c}}\right)\,, \label{JPsi:projector} \eqa
where $\varepsilon^{\mu}_{J/\psi}$ is the polarization vector of
$J/\psi$ satisfying $\varepsilon_{J/\psi}(\lambda)\cdot
\varepsilon_{J/\psi}^*(\lambda^\prime)
=-\delta^{\lambda\lambda^\prime}$ and $P\cdot \varepsilon=0$.
$N_c=3$, and ${\bf 1}_c$ stands for the unit color matrix.
The nonperturbative parameters, $\psi_{\eta_b}(0)$ and
$\psi_{J/\psi}(0)$,  are Schr\"{o}dinger wave functions at the
origin for $\eta_b$ and $J/\psi$,  which can be either
inferred from phenomenological potential models or
directly extracted from experiments.
By writing (\ref{Etab:projector}) and (\ref{JPsi:projector})
the way as they are,  it is understood that
$M_{\eta_b}=2m_b$ and $M_{J/\psi}=2m_c$ have been assumed.

We commence with the strong decay amplitude.  As can be seen in
Fig.~\ref{Feynman:Diag}, the lowest order QCD-initiated
contribution starts already at one loop.  Note that the photon can
only be emitted form the $c$ quark line, because the even
$C$-parity of $\eta_b$ forbids the $\gamma$ to attach to its
constituents. Using the projection operators in
(\ref{Etab:projector}) and (\ref{JPsi:projector}), it is
straightforward to write down the Feynman rules for the strong
decay amplitude:
\bqa {\cal M}_{\rm str}&=& -C_{\rm str}\,e_c e \,g_s^4\,
{\psi_{\eta_b}(0)\,\psi_{J/\psi}(0) \over 8 \sqrt{m_b m_c}}
\int\!\! {d^4k_1 \over (2\pi)^4}\,{1\over k_1^2}\, {1\over k_2^2}
\nn \\
& \times & \left[ { {\rm
tr}[(\not\!{Q}+2m_b)\gamma_5\gamma_\nu(\not \! p_b -\not\!
k_1+m_b)\gamma_\mu] \over (p_b-k_1)^2-m_b^2 } +
(\mu\leftrightarrow \nu,\;k_1\leftrightarrow k_2) \right ]
\nn\\
& \times & \left\{ {{\rm tr}[\not\! \varepsilon^*_{J/\psi} (\not\!
P+ 2m_c) \gamma^\mu(\not \! p_c -\not\! k_1+m_c)\not\!
\varepsilon^*_\gamma (\not\! k_{2} -{\not \!
p_{\bar{c}}}+m_c)\gamma^\nu] \over
((p_c-k_1)^2-m_c^2)\,((p_{\bar{c}}-k_2)^2-m_c^2)} \right.
\nn\\
&+& {{\rm{tr}}[\not\! \varepsilon^*_{J/\psi}(\not\! P+ 2m_c)
\not\! \varepsilon^*_\gamma (\not \!p_c +\not\! k
+m_c)\gamma^\mu(\not\! k_2 -{\not\! p_{\bar c}}+m_c)\gamma^\nu]
\over ((p_c+k)^2-m_c^2)\,((p_{\bar{c}}-k_2)^2-m_c^2)}
\nn\\
&+ & \left. { {\rm tr} [\not\! \varepsilon^*_{J/\psi} (\not\!
P+2m_c) \gamma^\mu(\not\! p_c -\not \! k_{1}+m_c)\gamma^\nu(-
\not\! p_{\bar c} -\not\! k+m_c)\not\! \varepsilon^*_\gamma  ]
\over ((p_c-k_1)^2-m_c^2)\,((p_{\bar{c}}+k)^2-m_c^2)} \right\}\,,
\label{QCD:ampl:Feyn:Rule}
\eqa
where the corresponding color factor $C_{\rm str}=N_c^{-1}\,{\rm
tr}(T^a T^b){\rm tr}(T^a T^b)={2\over3}$.  The  momenta carried by
two internal gluons are labeled by $k_1$, $k_2$, respectively,
which are subject to the constraint $k_1+k_2=Q$.  Notice that the
box diagrams in Fig.~\ref{Feynman:Diag} are related to one-loop
four- or five-point functions, and the corresponding loop
integrals are ultraviolet finite. Moreover, the occurrence of
heavy $b$ and $c$ masses ameliorate the infrared behavior of the
loop integral so that the result turns out to be simultaneously
infrared finite. Since there is no need for regularization, we
have directly put the spacetime dimension to four.

After completing the Dirac trace in (\ref{QCD:ampl:Feyn:Rule}), we
end up with terms which do not immediately possess the desired
tensor structure of (\ref{Lorentz:tensor:structure}), instead with
one index of Levi-Civita tensor contracted to the loop momentum
variable. Of course, when everything is finally worked out, all
these terms must conspire to arrive at the desired Lorentz
structure.  Conversely, one may exploit this knowledge to simplify
the algebra prior to performing the loop
integral~\cite{Guberina:1980xb,Korner:1982vg}. First pull out the
partial amplitude $M_{\mu\nu}$ through ${\cal M}_{\rm
str}=M_{\mu\nu}\,\varepsilon^{*\mu}_{J/\psi}\,
\varepsilon^{*\nu}_{\gamma}$. Eq.~(\ref{Lorentz:tensor:structure})
then demands
\bqa M_{\mu\nu} &=& {\cal A}_{\rm str} \:\epsilon_{\mu \nu \alpha
\beta}\, Q^\alpha\,k^\beta\,, \label{red:amp:mu:nu} \eqa
which is compatible with the expectation that $M_{\mu\nu}$ would
vanish unless $\mu$, $\nu$ are in transverse directions.
Contracting both sides of (\ref{red:amp:mu:nu}) with
$\epsilon^{\mu \nu \rho\sigma} Q_\rho k_\sigma$, we can express
the reduced amplitude as
\bqa
{\cal A}_{\rm str} &=& {1\over 2\,(k\cdot Q)^2}\, M_{\mu\nu}\,
\epsilon^{\mu \nu \rho\sigma} \,Q_\rho\,k_\sigma\,.
\eqa

After this manipulation is done,  it is convenient to adopt a new
loop momentum variable $q$, which is related to the old one
via $k_1=(Q+q)/2$ and $k_2=(Q-q)/2$. We end in a concise
expression:
\bqa {\cal A}_{\rm str} &=& -{e_c e \,g_s^4 \over 6\,\pi^2}
\sqrt{m_c\over m_b} \,{\psi_{\eta_b}(0)\,\psi_{J/\psi}(0)\over
(m_b^2-m_c^2)^2} \,  f\left({m_c^2\over m_b^2}\right)\,,
\label{red:am:str}
\eqa
where
\bqa
\label{integral}
f\left({m_c^2\over m_b^2}\right)={8\over
i\pi^2} \int\!d^4 q \, {(k\cdot Q)\,q^2-(k \cdot q)\,(Q\cdot
q)\over (q+Q)^2\,(q-Q)^2\, (q^2+2k\cdot q-P^2)\,(q^2-2k\cdot
q-P^2)}\, . \label{definition:f}
\eqa
Since $f$ is dimensionless, it can depend upon $m_b$ and $m_c$
only through their dimensionless combination. It is interesting to
note that the $b$ quark propagator has been canceled in this
expression. The loop integral can be performed analytically by the
standard method, and the result is
\bqa
{\rm Re}f(u)&=& {2(1-u)\over 2-u}\ln\left[{u\over
2(1-u)}\right]-{2\over 1+u} \left\{\ln^2 2+{1\over 2}\ln^2 u
+\ln[2-u]\ln\left[{u\over 2(1-u)}\right]
\right.
\nn\\
&+&\ln u\ln \left[{2\over 1-u}\right] -u\ln \left[{u\over
2-u}\right] \ln\left[{u\over 2(1-u)}\right]+2\,{\rm Li}_2[-u]+
{\rm Li}_2\left[{u-1\over 2u}\right]
\nn \\
&+& \left. 2\,{\rm Li}_2\left[{u\over 2}\right] -{\rm
Li}_2\left[{u^2-u\over 2}\right]-u\,{\rm Li}_2\left[2-{2\over
u}\right]\right\}\,,
\label{Ref:analyt}
\\
{\rm Im} f(u)&=&2\pi \left\{{1-u\over 2-u}+{u\ln u\over
1+u}-\ln[2-u] \right\}\,,
\label{Imf:analyt}
 \eqa
where ${\rm Li}_2$ denotes the dilogarithm (Spence) function. We
will expound the derivation of this result in the Appendix. Note
$f$ has an absorptive part, which reflects that the intermediate
gluons are kinematically permissible to stay on shell. One can
apply Cutkosky's cutting rule to verify (\ref{Imf:analyt}). The
real and imaginary parts of $f$ as function of $m_c^2/m_b^2$ are
displayed in Fig.~2.

\begin{figure}[tb]
\begin{center}
\includegraphics[scale=0.5]{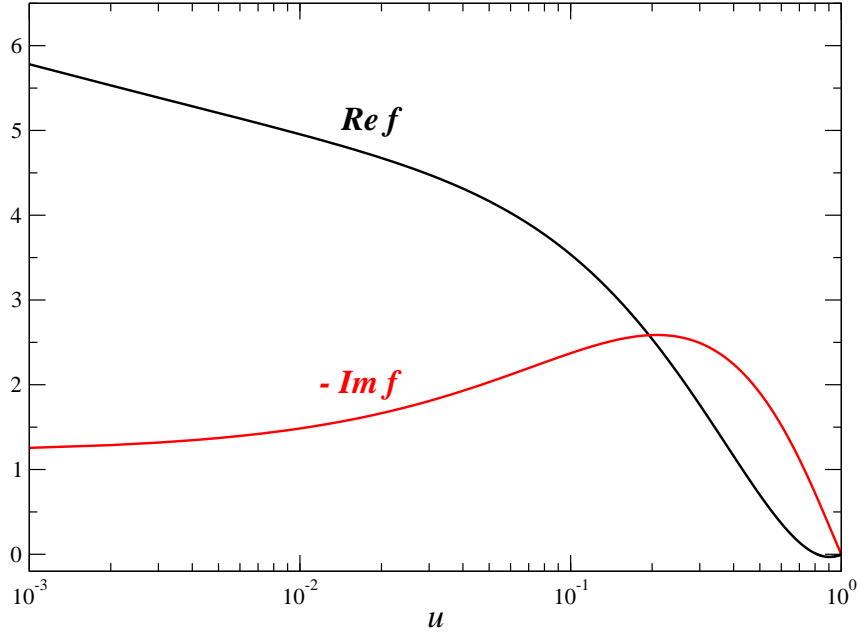}
\caption{Real and imaginary parts of $f(u)$.}
\end{center}
\label{shape:of:f}
\end{figure}

It is instructive to know the asymptotic behavior of
$f$ in the $u\to 0$ limit, which can be readily
read out from (\ref{Ref:analyt}) and (\ref{Imf:analyt}):
\bqa {\rm Re}f(u)&\longrightarrow& (1- 2\,\ln 2)\,\ln u +
{\pi^2\over 3}-\ln2+\ln^2 2\,,
\\
{\rm Im}f(u)&\longrightarrow& \pi\,(1- 2\,\ln 2)\,. \eqa
Note as $u$ approaches 0,
the imaginary part remains finite, whereas the real part blows up
logarithmically. This trend can be clearly
visualized in Fig.~2.  The logarithmic divergence in
the $m_c\to 0$ limit is obviously of infrared origin.
Nevertheless, this does not pose any practical problem,
since a non-relativistic description for a zero-mass bound state,
as well as the resulting predictions, should not be trusted anyway.
It is interesting to note that,  the analogous radiative decay process,
$\Upsilon\to \eta_c\gamma$, has a different asymptotic behavior,
in which both the real and imaginary parts of the QCD amplitude
admits a finite limit~\cite{Guberina:1980xb}.

We next turn to the pure QED contribution to this radiative decay
process. One may naively expect that this contribution is much
less important than the QCD contribution. However, it turns out
that this expectation is not true and the QED contribution  must
be retained. Due to the neutral color charge of photon, this
radiative decay can arise at tree level from the photon
fragmentation, as shown in Fig.~\ref{Feynman:Diag}~(4). Obviously,
the fragmentation type contribution is much more dominant over other
types of QED diagrams.  To calculate this contribution, a
necessary input is the $\gamma-J/\psi$ coupling, which is
characterized by the $J/\psi$ decay constant\footnote{ In
conformity with the sign convention adopted in
(\ref{JPsi:projector}), a minus sign is compulsory to put in here,
which takes into account the Grassmann nature of the quark field
operator.}:
\bqa
\langle J/\psi(\lambda)|\overline{c}\gamma^\mu c|0\rangle &=&
-g_{J/\psi} \, \epsilon^{*\mu}(\lambda)\,. \eqa
In the non-relativistic limit, $g_{J/\psi}$ is linked to
$\psi_{J/\psi}(0)$ through the relation $g_{J/\psi}=2^{3/2}
N_c^{1/2} \,m_c^{1/2}\,\psi_{J/\psi}(0)$, which can be derived
from (\ref{JPsi:projector}). The calculation is much easier than
its QCD counterpart,  and the reduced QED amplitude is
\bqa {\cal A}_{\rm em} &=& e_c e_b^2\,e^3\,N_c\, \sqrt{m_c\over
m_b} \: {\psi_{\eta_b}(0)\,\psi_{J/\psi}(0)\over
m_c^2\,(m_b^2-m_c^2)}\,. \label{red:am:em} \eqa

Substituting (\ref{red:am:str}) and (\ref{red:am:em}) into the
formula
\bqa \Gamma[\eta_b \to J/\psi\gamma ] &=& { |{\bf k}|^3 \over
4\,\pi}\,|{\cal A}_{\rm str}+{\cal A}_{\rm em}|^2\,, \eqa
we then obtain the desired partial width. Here
$|{\bf k}|=(m_b^2-m_c^2)/m_b$  is the photon momentum
in the $\eta_b$ rest frame. This formula already takes
into account the sum over transverse
polarizations of both $J/\psi$ and $\gamma$.

For phenomenological purpose,  it is instead more convenient to
have an expression for the branching ratio, where
$\psi_{\eta_b}(0)$ drops out:
 \bqa {\rm Br}[\eta_b\to J/\psi\gamma]&=&{8\,
e_c^2\, \alpha\, \alpha_s^2 \over 3\,\pi}
\,{m_c\,\psi^2_{J/\psi}(0)\over m_b^2 \,(m_b^2-m_c^2) }
\,\left|f\left({m_c^2\over m_b^2} \right)-g\left({m_c^2\over
m_b^2}\right)\right|^2\,, \label{Branch:Ratio} \eqa
where
\bqa g(u)&=& {9\,\pi\, e_b^2 \,\alpha\over 2\, \alpha_s^2}\,
{1-u\over u } \eqa
encodes the electromagnetic contribution.
In deriving this, we have approximated the total width of
$\eta_b$ by its gluonic width:
\bqa
\Gamma_{\rm tot}[\eta_b]\approx \Gamma[\eta_b\to g g]
&= & {8
\pi \,\alpha_s^2 \over 3\, m_b^2} \,\psi_{\eta_b}^2(0)\,,
\label{etab:tot:width}
\eqa
where the LO expression in $\alpha_s$ and $v_b$ is used  for
simplicity.

Eq.~(\ref{Branch:Ratio}) constitutes the key formula of this work.
This equation conveys that, despite the adversity caused by the
suppression $\alpha/\alpha_s^2$, the QED fragmentation
contribution nevertheless enjoys the kinematic enhancement of
$m_b^2/m_c^2$ relative to the QCD amplitude. For the physical
masses of $b$ and $c$, these two competing effects have comparable
magnitudes.  In the asymptotic regime as $m_b/m_c\gg 1$, the QED
amplitude will eventually dominate over its  QCD  counterpart,
because the enhancement factor $m_b^2/m_c^2$ of the former is much
more eminent than the mild $\log(m_b^2/m_c^2)$ rising of the
latter. At any rate, it is imperative to include the QED
fragmentation contribution.  Moreover, it is important to
recognize that the interference between these two amplitudes is
{\it destructive}.  This is opposite to what occurs in the
analogous decay process $\Upsilon\to
\eta_c\gamma$~\cite{Guberina:1980xb} and in the decay $\eta_b\to
J/\psi\,J/\psi$~\cite{Jia:2006rx}, where the interference is {\it
constructive}.

\section{Phenomenology}
\label{phenomenology}

\subsection{Observation Potential of $\eta_b\to J/\psi\gamma$ in Tevatron
and LHC}

It is now the time to explore the phenomenological implication of
(\ref{Branch:Ratio}). The input parameters are $m_b$, $m_c$,
$\alpha$, $\alpha_s$ and $\psi_{J/\psi}(0)$, all of which can be
inferred from other independent sources. The wave function at the
origin for $J/\psi$ can be extracted from its electronic width:
\bqa
\Gamma[J/\psi\to e^+e^-] &=&
{4 \pi \,e_c^2 \,\alpha^2 \over m_c^2} \psi_{J/\psi}^2(0)\,,
\label{onium:ee:width}
\eqa
where the LO formula in $\alpha_s$ and $v_c^2$ is used for simplicity.
Using the measured dielectron width 5.55 keV~\cite{Yao:2006px}, we
obtain $\psi_{J/\psi}(0)=0.205$ ${\rm GeV}^{3/2}$ for $m_c=1.5$
GeV. Taking  $m_b =M_{\eta_b}/2 \approx 4.7$ GeV,
$m_c=1.5$ GeV, $\alpha=1/137$ and $\alpha_s(m_b)=0.22$,
we then find
\bqa
{\rm Br}[\eta_b\to J/\psi\, \gamma] &= & (1.5 \pm 0.8) \times
10^{-7}\,.
\label{BR:EJP:predct}
\eqa
The uncertainty is estimated by varying $\alpha_s(\mu)$ between
0.18 and 0.26 (which corresponds to slide the scale from
$\mu=2m_b$ to $2m_c$), as well as taking into account the errors
in the measured $\Gamma_{e^+e^-}$ (of $\pm 0.14$ keV). The
destructive interference between electromagnetic and strong
amplitudes has pronounced effect. For the central values of input
parameters, omitting the QED contribution will result in a
prediction of $3.5\times 10^{-7}$, which is more than twice larger
than the actual value.  To develop a concrete perception, we
enumerate the values of $f$ and $g$ evaluated at
$u=m_c^2/m_b^2=0.10$ and $\alpha_s=0.22$:
\begin{equation} f= 3.5 - 2.4\,i\,,\hspace{1.3 cm} g=2.1\,.
\end{equation}
Clearly, ${\rm Re }f$, ${\rm Im }f$ and $g$ all have comparable
magnitudes. As a result, the destructive interference effect is
particularly important.

The numerical prediction presented in (\ref{BR:EJP:predct}) is
obtained by only using the tree-level matching coefficients for
the total $\eta_b$ width and leptonic width of $J/\psi$. One may
worry that this oversimplified procedure will induce some error
because it is known that the next-to-leading perturbative
corrections to both quantities, especially to the $J/\psi$
leptonic width, are large. Let us assess their effects now. To the
NLO accuracy in $\alpha_s$, one needs to multiply
Eq.~(\ref{etab:tot:width}) by $1+(53/2-31\,\pi^2/ 24-8\, n_f /9)\,
\alpha_s(2m_b)/\pi$ ($n_f$ stands for the number of active light
flavors), as well as multiply Eq.~(\ref{onium:ee:width}) by
$(1-8\alpha_s(2m_c)/ 3\pi)^2$~\cite{Bodwin:1994jh}. Incorporating
these corrections amounts to multiplying (\ref{BR:EJP:predct}) by a
factor~\footnote{It should be kept in mind that the relativistic
correction to leptonic decay of $J/\psi$ is also large. We have
not considered this complication.}
\bqa
\left(1+{10.2\,\alpha_s(2m_b)\over \pi}\right)^{-1}
 \left(1-{8\alpha_s(2m_c)\over 3\pi}\right)^{-2}
 &=& 1.04\,,
 \eqa
where $n_f=4$ has been taken.  Since including the perturbative
radiative corrections has negligible net impact,  we will keep
using (\ref{BR:EJP:predct}) in the following phenomenological
analysis.

It is enlightening to compare (\ref{BR:EJP:predct}) with the
NRQCD prediction to the branching ratio for $\eta_b$
decays to double $J/\psi$~\cite{Jia:2006rx}:
\bqa {\rm Br}[\eta_b\to J/\psi\, J/\psi] &= &
2.4^{+4.2}_{-1.9}\times 10^{-8}\,.
\label{etab:to:JpsiJpsi:prediction}
\eqa
Notice that the branching ratio of our radiative decay process is
almost one order-of-magnitude larger than that of this hadronic
decay process! The reason can be traced as follows. The double
$J/\psi$ decay mode, though being a hadronic one, has maximally
violated the helicity selection rule of pQCD. As a result, the
branching ratio gets severely suppressed, $\propto \alpha_s^2
\,v_c^{10} (m_c/ m_b)^8$~\cite{Jia:2006rx}.  In contrast, if we
count $f\sim {\cal O}(1)$~\footnote{The slow rising
$\ln(m_b^2/m_c^2)$ term in ${\rm Re} f$ is of no concern here for
physical $b$ and $c$ masses. Also we neglect the pure QED
contribution for the lucidity of the argument.},
Eq.~(\ref{Branch:Ratio}) then implies that our radiative decay
process admits a scaling behavior ${\rm Br}\sim \alpha
\,\alpha_s^2 \,v_c^3 \,(m_c/ m_b)^4$, which is much more mildly
suppressed by powers of $1/m_b^2$ and $v_c$ relative to $\eta_b\to
J/\psi J/\psi$, hence is more favorable.

Experimentally $J/\psi$ can be cleanly reconstructed  by decays to
lepton pairs.  Multiplying (\ref{BR:EJP:predct}) by the branching
ratios of 12\% for $J/\psi$ decays to $\mu^+ \mu^-$ and $e^+e^-$,
we obtain $ {\rm Br}[\eta_b\to J/\psi \gamma \to l^+ l^- \gamma] =
(0.8-2.8)\times 10^{-8}$. The total cross section for producing
$\eta_b$ at Tevatron has been estimated to be about 2.5 $\mu{\rm
b}$~\cite{Maltoni:2004hv}. The production cross section for the $
l^+ l^- \gamma$ events is thus about $0.02-0.07$ pb. For the full
Run 1 data of 100 ${\rm pb}^{-1}$, we then obtain between 2 and 7
produced events. Because the kinematical cuts, as well as taking
into account the acceptance and efficiency for detecting leptons,
will further cut down this number, it seems not so fruitful to
assiduously seek the $\eta_b$ through this radiative decay mode in
Run 1 data sample.

Tevatron Run 2 aims to achieve an integrated luminosity of 8.5
${\rm fb}^{-1}$ by 2009. Assuming equal $\sigma(p\overline{p}\to
\eta_b +X)$ at $\sqrt{s}=1.96$ and $1.8$ TeV, we then estimate
there are about 200-600 produced events. The product of acceptance
and efficiency for detecting $J/\psi$ decay to muon pair is
estimated to be $\epsilon\approx 0.1$~\cite{Braaten:2000cm}. It
may sound reasonable to assume the corresponding factor for the
electron pair also of the same magnitude. Multiplying the number
of the produced events by $\epsilon$, we expect between 20 and 60
observed events in the full Run 2 period.  This is quite
encouraging, but we should be cautious about the fact that the
major QCD background events, the associated $J/\psi+\gamma$
production, could also be copious.  At the Tevatron and LHC
colliders, the dominant production mechanism for these events is
through $gg$ fusion~\cite{Roy:1994vb,Kim:1996bb,Mathews:1999ye}. A
naive analysis indicates that the direct $J/\psi+\gamma$
production with an invariant mass near $9.4$ GeV preponderates
over that from the $\eta_b$ decays.

It is important to keep in mind that the $J/\psi$ stemming from
the radiative $\eta_b$ decay must be transversely polarized. This
characteristic may be used to discriminate the signal events from
the background events, because $J/\psi$ from the latter processes
can also be longitudinally polarized. Furthermore, the detailed
kinematical distributions of background processes need to be
thoroughly studied, in order to guide experimentalists to choose
the optimal kinematical cuts to oppress as many background as
possible, and in the meanwhile without significantly sacrificing
the signal events. This is somewhat beyond the scope of this work
and needs further independent studies.

The forthcoming LHC experiments will greatly increase the number
of the produced $\eta_b\to l^+ l^- \gamma$ events.  To assess the
discovery potential of this mode at LHC, we need first know the
inclusive production rate for $\eta_b$.  There are rough estimates
for the $\chi_{b0,2}$ cross sections at LHC, which are about 6
times larger than the corresponding ones for producing them at
Tevatron~\cite{Braguta:2005gw}. Assuming the same scaling also
holds for $\eta_b$, we then obtain the cross section for $\eta_b$
at LHC to be about 15 ${\rm \mu b}$,  subsequently the production
cross section for the $\l^+ \l^- \gamma$ events to be about
0.1-0.4 pb.  For 300 ${\rm fb}^{-1}$ data, which is expected to be
collected in one year run at LHC design luminosity,  the number of
produced events may reach about $3\times 10^4-1\times 10^5$.
Multiplying the number of the produced events by $\epsilon$, we
expect between $3\times 10^3$ and $1\times 10^4$ observed events
per year.

Based on this analysis,  we are tempted to conclude that, the
chance of observing $\eta_b$ at LHC through this mode is very
promising. With such a large amount of signal events, it is
possible to measure the leptonic angular distribution to pin down
the polarization of $J/\psi$.  As has been stressed, in order to
effectively selected the signal events out of the abundant
background events, one needs to develop a thorough understanding
towards the QCD background.

\subsection{Radiative decay of $\eta_b$ ($\eta_c$) into $\phi$}

When coping with light mesons in hard exclusive processes, the
most natural treatment for them is using light-cone wave
functions. Indeed, heavy quarkonium radiative decays to light
mesons,  such as $\Upsilon \to
f_2(1270)\gamma$~\cite{Ma:2001tt,Fleming:2004hc} 
(Ref.~\cite{Fleming:2004hc} has heavily exploited the EFT machinery, 
{\it i.e.} NRQCD combined with SCET), $J/\psi(\Upsilon) \to \eta
(\eta^\prime)\gamma$~\cite{Yang:2004wy,Li:2005ug}, have been
studied along this line. On the other hand, the constituent quark
model,  which treats the light meson as a non-relativistic bound
state,  is also frequently invoked as an alternative method for a
quick order-of-magnitude estimate.  Various radiative decay
processes, {\it e.g.} $J/\psi$ decays into light pseudo-scalar and
$P$-wave mesons,  as well as $\chi_{cJ}$ into light vector mesons,
have already been studied in this
context~\cite{Korner:1982vg,Gao:2006bc}.

We may apply the same strategy to analyze the radiative decay
$\eta_b (\eta_c)\to V\,\gamma$.  We take $\eta_b\to \phi\gamma$ as
a representative. By regarding $\phi$ as a strangenium, we can
directly use (\ref{Branch:Ratio}), only with some trivial changes
of input parameters. We take the $m_s = M_\phi/2 \approx 0.5$ GeV.
The wave function at the origin of $\phi$, $\psi_{\phi}(0)$, can
be extracted  through (\ref{onium:ee:width}) from its measured
electronic width of $1.27\pm 0.04$ keV~\cite{Yao:2006px}.  Taking
$m_b=4.7$ GeV, varying the strong coupling constant between
$\alpha_s(2m_b)=0.18$ and $\alpha_s(2m_s)= 0.51$, and including
the experimental uncertainty in $\Gamma_{e^+e^-}$, we obtain
\bqa
{\rm Br}[\eta_b\to \phi\gamma] &= & (0.3-6.9)\times
10^{-8}\,.
\eqa
The QED contribution dominates in this case due to the larger
ratio of $m_b$ to $m_s$.  To see this concretely, we list the
values of $f$ and $g$ evaluated at $u=m_s^2/m_b^2\approx 0.01$ and
$\alpha_s=0.22$:
\bqa f &=& 4.9 - 1.5\,i\,,\hspace{1.3 cm} g=23.5 \,. \eqa
The dominance of $g$ over $|f|$ is apparent.  Neglecting QED
contribution for this $\alpha_s$ value will decrease the branching
fraction by one order of magnitude.  For the absence of clean
signature for $\phi$,  such a rare decay mode will be extremely
difficult to observe in hadron collision experiments such as LHC.
It may be interesting to compare this radiative decay channel with
the following hadronic mode~\cite{Jia:2006rx}:
\bqa {\rm Br}[\eta_b\to \phi\phi] &= & (0.9-1.4)\times 10^{-9}\,.
\eqa
This $2\phi$ mode is even smaller, and hopeless to be seen
experimentally.

One can proceed to consider the analogous decay process
$\eta_c \to \phi\gamma$.  Parallel to the preceding analysis,
taking $m_c=1.5$ GeV, and varying $\alpha_s(\mu)$ from
$\alpha_s(2 m_c)=0.26$ to
$\alpha_s(2m_s)=0.51$,   we obtain
\bqa {\rm Br}[\eta_c\to \phi\gamma] &= & (2.1-8.6)\times
10^{-7}\,. \label{etac:phi:phi:pQCD} \eqa
The destructive interference pattern is similar to that in
$\eta_b\to J/\psi\gamma$.  This decay mode is still too much
suppressed to be observed in the functioning charmonium factory
like BES~II, perhaps also in the forthcoming BES III experiment.
It seems also rather difficult to observe this decay mode in the
current and future hadron collision facilities.

\section{Summary}
\label{summary}

The motif of this work is to suggest one viable way to ferret out
$\eta_b$ in the functioning and forthcoming hadron collider
facilities, that is, through its radiative decay into $J/\psi$.
This decay mode owns the advantage that both $J/\psi$ and photon
can be tagged cleanly. The presence of $J/\psi$ is particularly
helpful to reduce the combinatorial background. In this regard,
this mode is practically much more useful than the purely
electromagnetic decay $\eta_b\to \gamma\gamma$.

By an explicit pQCD calculation based on NRQCD approach, we infer
the branching ratio of this process to be of order $10^{-7}$.
Although the absolute value of this ratio is small, it is already
larger than that of the hadronic decay mode $\eta_b\to J/\psi
J/\psi$,  which was previously thought of as a golden mode for
searching $\eta_b$. Our analysis indicates that the chance of
observing $\eta_b$ through this decay channel, followed by
$J/\psi$ decay to a lepton pair, with a corresponding branching
ratio about $10^{-8}$, seems still open in Tevatron Run 2, and is
quite promising in the forthcoming LHC experiment.  However, one
should bear in mind that the major QCD background events are
expected to greatly outnumber the desired signal events. A
thorough study of the associated $J/\psi+\gamma$ production is
welcome,  in order to help experimentalists to impose optimal
kinematical cuts to singlet out the signal events from the
abundant background events. The transverse polarization of
$J/\psi$ in the signal events should be employed to effectively
veto the background.

While this paper is being written,  we are informed of the same
$\eta_b \to J/\psi \gamma$ process being also considered by Gao,
Zhang and Chao~\cite{Gao:2006}. These authors evaluated
loop integrals numerically. It has been checked that once
the same input parameters are assumed, our numerical prediction for
the branching ratio is compatible with theirs.
These authors have also studied various other radiative
decay channels of bottomonium to charmonium.

\acknowledgments
The work of  G.~H., C.~F.~Q. and P.~S. was supported in part
by the Natural Science Foundation of China and by the
Scientific Research Fund of
GUCAS (NO. 055101BM03). Y.~J. is supported in part by
NSFC under Grant No.~10605031 (effective since January 2007).

\appendix
\section{Deriving analytical expression for $f$}\label{f:analytic}

In this Appendix we illustrate how to reduce the one-loop
four-point function in (\ref{definition:f}) to the sum of
much simpler two- and three-point scalar integrals.
We start with the dimensionless integral $f$:
\begin{equation}
f\left({m_c^2\over m_b^2}\right)={8\over i\pi^2} \int\!d^4 q
{(k\cdot Q)\,q^2-(k \cdot q)\,(Q\cdot q)\over (q+Q)^2\,(q-Q)^2\,
(q^2+2k\cdot q-P^2)\,(q^2-2k\cdot q-P^2)}\,,
\label{def:f}
\end{equation}
where $Q^2=4m_b^2$ and $P^2=4m_c^2$. The $+i\varepsilon$
prescription in the propagators is implicitly implied.

First note the second term in the integrand can be simplified by
using the identity of fractional sum:
\bqa && \int\!d^4 q { (k \cdot q)\,(Q\cdot q)\over
(q+Q)^2\,(q-Q)^2\, (q^2+2k\cdot q-P^2)\,(q^2-2k\cdot q-P^2)}
\nn \\
&=& {1 \over 16}\int\!d^4 q\left[{1\over(q-Q)^2} - {1\over
(q+Q)^2}\right]\left[{1\over q^2-2k\cdot q-P^2}-{1\over
q^2+2k\cdot q-P^2} \right] \nn
\\
&=& {1 \over 8}\int\! d^4q \,{1\over (q-Q)^2}\left[{1\over
q^2-2k\cdot q-P^2}-{1\over q^2+2k\cdot q-P^2} \right]\,,
\label{simpl:f1}
 \eqa
which is nothing but the scalar 2-point functions,
thus can be trivially worked out.

Disentangling the first term in (\ref{def:f}) needs slightly more
labor. Since the integrand is an even function of $q$,
we are free to add terms linear in $q$ in the numerator,
without influencing the result:
\bqa
&& \int\!d^4 q \, { q^2 \over (q+Q)^2\,(q-Q)^2\, (q^2+2k\cdot
q-P^2)\,(q^2-2k\cdot q-P^2)}
\nn \\
&=& {1 \over Q^2+P^2}\int\!d^4 q \, {(Q^2+P^2)\,q^2+2Q^2 \,k\cdot
q +2P^2\, Q\cdot q +Q^2 P^2-Q^2P^2 \over (q+Q)^2\,(q-Q)^2\,
(q^2+2k\cdot q-P^2)\,(q^2-2k\cdot q-P^2)}\nn
\\
&=&{Q^2\over Q^2+P^2}\int\! d^4q \, {1 \over (q+Q)^2\, (q-Q)^2\,
(q^2-2k\cdot q-P^2)}
\nn\\
& + & {P^2\over Q^2+P^2}\int\! d^4q \, {1 \over
(q-Q)^2\,(q^2+2k\cdot q-P^2) (q^2-2k\cdot q-P^2)}\,,
\label{simpl:f2:f3}
 \eqa
which consists of two independent 3-point scalar integrals.
They can also be worked out in closed form,
following the method outlined in Ref.~\cite{'tHooft:1978xw}.

With the aid of (\ref{simpl:f1}) and (\ref{simpl:f2:f3}), we can
decompose the original $f$ into three pieces:
\bqa
f(u)&=& f_1(u)+f_2(u)+f_3(u)\,,
\eqa
where $u\equiv P^2/Q^2=m_c^2/m_b^2$, and
\bqa
f_1(u) &=& {i \over \pi^2}\int\! d^4q \,{1\over
(q-Q)^2}\left[{1\over q^2-2k\cdot q-P^2}-{1\over q^2+2k\cdot
q-P^2} \right]
\nn \\
 &=&
 {2\,(1-u)\over 2-u}\,\ln\left[{u \over 2(1-u)}\right]
+ 2\pi\, i \left({1-u\over 2-u}\right)\,,
\\
f_2(u) &=&{8\,Q^2\,k\cdot Q \over i\pi^2(Q^2+P^2)}\int\! d^4q \,
{1 \over (q+Q)^2\, (q-Q)^2\, (q^2-2k\cdot q-P^2)}
\nn\\
&=& - \,{2\over 1+u} \, \left\{\ln^2 2+{1\over
2}\ln^2u+\ln[2-u]\,\ln\left[{u\over 2(1-u)}\right]+\ln
u\,\ln\left[{2\over 1-u}\right] \right.
\nn\\
&+& \left. 2\,{\rm Li}_2[-u]+{\rm Li}_2\left[{u-1\over 2u}\right]
+2\, {\rm Li}_2\left[{u\over 2}\right]-{\rm Li}_2\left[{u^2-u
\over
2}\right]\right\}
- {2\pi\, i \over 1+u}\,\ln[2-u] \, ,
\\
f_3(u) &=&{8\,P^2\,k\cdot Q \over i\pi^2(Q^2+P^2)}\int\! d^4q \,
{1 \over (q-Q)^2\,(q^2+2k\cdot q-P^2) (q^2-2k\cdot q-P^2)} \nn \\
&=& {2\,u\over 1+u}\left\{\ln\left[{u\over
2-u}\right]\ln\left[{u\over 2(1-u)}\right]+{\rm
Li}_2\left[2-{2\over u}\right]\right\}
\nn \\
&+& \,2\pi i \left({u\over 1+u}\right) \ln\left[{u\over
2-u}\right]\,.
\eqa
One can then readily reproduce the analytic results shown
in (\ref{Ref:analyt}) and (\ref{Imf:analyt}).


\end{document}